\documentclass[aps,prd,reprint,amsmath,amssymb,showpacs,superscriptaddress,nofootinbib]{revtex4-1}
\usepackage{epsfig,slashed,color,bm,natbib,comment,enumitem,ulem,cancel}
\def\vect#1{\mbox{\boldmath $#1$}}
\def\hvect#1{{\hat{\mbox{\boldmath $#1$}}}}
\def\matri#1{{\mbox{\sf  #1}}}

\def\nd#1#2{{\frac{d #1}{d #2}}}
\def\hnd#1#2#3{{\frac{d^{#3} #1}{d #2 ^{#3}}}}

\def\tfrac#1#2{{\textstyle\frac{#1}{#2}}}

\newcommand{\vpsi}{\varphi}
\def\christoffel#1#2#3{{^0}{\Gamma^{#1}}_{#2 #3}}

\newcommand{\comma}{, }
\newcommand{\be}{\begin{equation}}
\newcommand{\ee}{\end{equation}}
\newcommand{\bea}{\begin{eqnarray}}
\newcommand{\eea}{\end{eqnarray}}
\newcommand{\myDel}[1]{{\color{red}\ifmmode\cancel{#1}\else\sout{#1}\fi}}
\newcommand{\myNew}[1]{{\color{black}#1}}

\begin{document}
\title{Weyl gauge theories of gravity do not predict a second clock effect}
\author{M.P.~\surname{Hobson}}
\email{mph@mrao.cam.ac.uk}
\affiliation{Astrophysics Group, Cavendish Laboratory, JJ Thomson Avenue,
Cambridge CB3 0HE\comma UK}
\author{A.N.~\surname{Lasenby}}
\email{a.n.lasenby@mrao.cam.ac.uk}
\affiliation{Astrophysics Group, Cavendish Laboratory, JJ Thomson Avenue,
Cambridge CB3 0HE\comma UK}
\affiliation{Kavli Institute for Cosmology, Madingley Road, Cambridge
  CB3 0HA, UK}
%\date{Received 16 January 2020; accepted ???; published online ???}

\begin{abstract}
We consider Weyl gauge theories of gravity (WGTs), which are invariant
both under local Poincar\'e transformations and local changes of
scale. Such theories may be interpreted as gauge theories in Minkowski
spacetime, but their gravitational interactions are most often
reinterpreted geometrically in terms of a Weyl--Cartan spacetime, in
which any matter fields then reside. Such a spacetime is a straightforward
generalisation of Weyl spacetime to include torsion. As first
suggested by Einstein, Weyl spacetime is believed to exhibit a so-called second
clock effect, which prevents the existence of experimentally
observed sharp spectral lines, since the rates of (atomic) clocks
depend on their past history. The prevailing view in the
literature is that this rules out WGTs as unphysical. Contrary to this
viewpoint, we show that if one adopts the natural covariant derivative
identified in the geometric interpretation of WGTs, properly takes
into account the scaling dimension of physical quantities, and
recognises that Einstein's original objection requires the presence of
massive matter fields to represent atoms, observers and clocks, then
WGTs do not predict a second clock effect.
\end{abstract}

%Modified theories of gravity, Gauge field theories, 
%Conformal field theory, algebraic structures
%11.30.Cp Lorentz and Poincaré invariance

\maketitle
%%%%%%%%%%%%%%%%%%%%%%%%%%%%%%%%%%%%%%%%%%%%%%%%%%%%%%%%%%%%%%%%%%%%%%%%%%

\section{Introduction}
\label{sec:intro}

In 1918, Weyl proposed a unified theory of gravity and
electromagnetism \cite{Weyl18}, which was based on a generalisation of the
Riemannian spacetime geometry assumed in Einstein's theory of general
relativity. In particular, in Weyl's spacetime the
principle of relativity applies not only to the choice of reference
frames, but also to the choice of local standards of length. This
invariance under local changes of the unit of length (gauge) was
realized by the introduction of an additional `compensating' vector
field, that we shall denote by $B_\mu$, which Weyl attempted to
interpret as the electromagnetic 4-potential.

In spite of the elegance and beauty of Weyl's theory, it did not
achieve its original goal. It was soon recognized as being unable to
accommodate well-known properties of electromagnetism, since the Weyl
potential $B_\mu$ is not coupled to the electric current, but to the
dilation current of matter. Indeed, one may easily show that $B_\mu$
interacts in the same manner with both particles and antiparticles,
contrary to all experimental evidence about electromagnetic
interactions. It was only later realised \cite{Weyl31} that electromagnetism
was related to localisation of invariance under change of
quantum-mechanical phase and, much later, that $B_\mu$ might instead be
interpreted as mediating an additional {\it gravitational} interaction,
within a theory of gravity that is locally scale-invariant.

The first objection to Weyl's theory was, however, made by Einstein in
a note published as an addendum to Weyl's original paper, and applies
irrespective of whether $B_\mu$ is interpreted as mediating the
electromagnetic or gravitational interaction. Einstein claimed that
Weyl's theory predicts a so-called `second clock effect', which is not
experimentally observed.  This phenomenon is in addition to the usual
`first clock effect', which also occurs both in special and general
relativity and has been experimentally verified to high precision. 

As is well known, the latter refers to the fact that if two identical
clocks, initially synchronised, coincident and at relative rest,
follow different (timelike) worldlines in spacetime before being
brought back together, they will in general measure different elapsed
(proper) time intervals. Nonetheless, provided the two clocks then
remain coincident, they will thereafter continue to `tick' at the {\it
  same} rate. By contrast, in Weyl spacetime, if the field strength
$H_{\mu\nu} \equiv 2\partial_{[\mu} B_{\nu]}$ of the Weyl potential
does not vanish throughout the spacetime interior of the two
clock worldlines during their separation, then the clocks in this
scenario will `tick' at {\it different} rates even after they are
re-united, which is known as the second clock effect (SCE).  An immediate
physical consequence is that the existence of sharp spectral lines
would not be possible in the presence of a non-zero field strength
$H_{\mu\nu}$, since the rate of atomic clocks, as measured by some
periodic physical process, would depend on their past history.

The original discussions of the SCE, which subsequently involved Weyl,
Einstein, Eddington and Pauli, amongst others
\cite{Pauli19,Eddington20,Eddington24,Schulman97,Goenner04}, were
based on the fact that in a Weyl spacetime, the `norms' of parallel
transported vectors change in a manner that depends on the path taken
(although the angle between two vectors remains the same; this changes
only in general affine spaces for which the metric and connection are
fully independent quantities).  It was then argued that the norm of a
timelike vector that is parallel transported along a timelike
worldline can represent the `tick' rate of a clock, which hence leads
to a SCE (equally, if the parallel transported vector is spacelike,
then one may physically interpret the effect as the length of a rod
being dependent on its past history, which is again contrary to
experimental evidence).

\newpage
The association of the clock rate with the norm of a
parallel-transported vector is not trivial, particularly given that
length is not a well-defined concept in Weyl's spacetime, but one may come
to the same conclusion by defining a physically sensible notion of
proper time along (timelike) worldlines in Weyl spacetime, which
generalises the concept of proper time used in Riemannian spacetimes 
\cite{Ehlers12, Perlick87, Avalos18}. By reconsidering the two-clock
thought experiment outlined above, and computing the elapsed proper
time measured by each clock between their reunion and some subsequent
event, one again concludes that a Weyl spacetime does indeed exhibit a
SCE, unless the Weyl potential can be expressed as the gradient of
some smooth scalar field $B_\mu = \partial_\mu\phi$; this corresponds
to a so-called Weyl integrable spacetime (WIST), in which the field
strength $H_{\mu\nu}$ vanishes identically.

In this paper, we reconsider the issue of the SCE in the context of
Weyl gauge theories of gravity (WGTs)
\cite{Bregman73,Charap74,Kasuya75,Blagojevic02}. 
These theories are derived by
gauging the Weyl group, where one begins with some Minkowski spacetime
matter action that is invariant under global Weyl transformations,
which consist of Poincar\'e tranformations and dilations, and then
demands that the action be invariant under local Weyl
transformations, where the  group parameters become independent
arbitrary functions of position. This requires the introduction of
gauge fields, which are interpreted as mediating gravitational
interactions.  Although WGTs are most naturally interpreted as gauge
field theories in Minkowski spacetime, it is usual for them to be
reinterpreted geometrically, whereby the gravitational interactions
are considered in terms of the geometry of a Weyl--Cartan spacetime,
in which any matter fields then reside \cite{Blagojevic02,eWGTpaper}. Weyl--Cartan spacetime is a
straightforward generalisation of Weyl spacetime to include non-zero
torsion and reduces to Weyl spacetime on imposing the properly
covariant condition that the torsion vanishes. Since, as we will
confirm, the presence of torsion is irrelevant to considerations of
the SCE, it has thus previously been argued that Einstein's objection
to Weyl spacetime rules out WGTs as unphysical, unless the Weyl
potential is pure gauge \cite{Wheeler98,Spencer11,Wheeler13,Wheeler18}.

Contrary to this prevailing view, we demonstrate that WGTs do not
require this condition in order to avoid the presence of the SCE. In
particular, we show that the geometric interpretation of WGTs leads to
the identification of the Weyl covariant derivative as the natural
derivative operator, which differs from the covariant derivative
usually assumed in Weyl--Cartan spacetimes when applied to quantities
having non-zero scaling dimension (or Weyl weight) $w$.  This is
especially important when differentiating the tangent vector
$u^\mu(\lambda) = dx^\mu/d\lambda$ along an observer's worldline,
which we show must have Weyl weight $w=-1$, rather than being
invariant ($w=0$) as is usually assumed. Finally, we point out that,
since Einstein's objection to Weyl's theory is based on the
observation of sharp spectral lines, one requires the presence of
matter fields to represent the atoms, observers and clocks; it is thus
meaningless to consider the SCE in an empty Weyl--Cartan geometry.
Moreover, such `ordinary' matter is most appropriately represented by a
massive Dirac field, but in order to obey local Weyl invariance this
field must acquire a mass dynamically through the introduction of a
scalar compensator field, which we show is key to defining an interval
of proper time as measured by a clock along an observer's
worldline. On taking these considerations into account, WGTs do not
predict a SCE, even when the Weyl potential is not pure gauge.

The outline of our argument is
  as follows. The geometric interpretation of WGTs identifies the
  (inverse) translational gauge field as the vierbein components
  ${e^a}_\mu$, which have
  Weyl weight $w=1$ and relate the orthonormal tetrad frame
  vectors $\hvect{e}_a(x)$ and the coordinate frame vectors
  $\vect{e}_\mu(x)$ at any point $x$ in a Weyl--Cartan spacetime. The
  vectors $\hvect{e}_a(x)$ constitute a local Lorentz frame at
  each point, which defines a family of ideal observers whose
  worldlines are the integral curves of the timelike unit vector field
  $\hat{\mathbf{e}}_0$.  
Along a given worldline, the three spacelike
unit vector fields $\hat{\mathbf{e}}_i$ $(i=1,2,3)$ specify the
spatial triad carried by the corresponding observer, which may be
thought of as defining the orthogonal spatial coordinate axes of a
local laboratory frame that is valid near the observer's
worldline. 
In general, the worldlines need not be time-like geodesics, and hence
observers may be accelerating.  For some test particle (or other
observer) moving along some timelike worldline $\mathcal{C}$ given by
$x^\mu=x^\mu(\lambda)$, where $\lambda$ is some arbitrary parameter,
the components of the tangent vector to this worldline, as measured by
one of the above observers, will be $u^a(\lambda) = {e^a}_\mu
u^\mu(\lambda)$, which are physically observable quantities in WGTs and so
should be invariant ($w=0$) under Weyl scale gauge
transformations. Since the vierbein ${e^a}_\mu$ has weight $w=1$, the
weight of the components $u^\mu(\lambda)$ must thus be
$w=-1$.\footnote{As we will discuss in Section~\ref{sec:WCST}, one may
  reach the same conclusion by demanding that the physical distance,
  as opposed to the coordinate distance, along the curve $\mathcal{C}$
  is traced out at the same rate before and after a Weyl scale gauge
  transformation.} The length of the tangent vector is then
invariant under Weyl scale gauge transformations. Moreover, by working
in terms of the Weyl covariant derivative, we show that one may always
find a parameterisation $\xi=\xi(\lambda)$ for which the length of the
tangent vector remains equal to unity under parallel transport along
its worldline (and so $u^\mu(\xi)=dx^\mu/d\xi$ may be interpreted as
the particle 4-velocity). Consequently, the original argument for
suggesting the existence of the SCE is removed. 

Since $d/d\xi$ still has weight $w=-1$, however, the parameter $\xi$
cannot be interpreted as the proper time of a particle moving along
the worldline.  To resolve this issue, we note that in order for WGTs
to include `ordinary' matter, which is usually modelled by a Dirac
field, one must introduce a scalar compensator field $\phi$ with Weyl
weight $w=-1$ and make the replacement $m\bar{\psi}\psi \to
\mu\phi\bar{\psi}\psi$ in the Dirac action, where $\mu$ is a
dimensionless parameter but $\mu\phi$ has the dimensions of mass in
natural units. In particular, the functioning of any form of practical
atomic clock is based on the spacing of the energy levels in atoms,
which is then characterised in the clock's local Lorentz frame by the
Rydberg energy $E_{\rm R} = \tfrac{1}{2}\mu\phi\alpha^2$ (in natural
units), where $\alpha$ is the (dimensionless) fine structure
constant. As a result, an interval of proper time measured by the
clock along its worldline is in fact given by $d\tau = \phi\,d\xi$,
which is invariant $(w=0)$ under Weyl scale gauge transformations, as
required for a physically observable quantity. Applying this proper
time definition to the two-clock thought experiment discussed above,
one finds immediately that the clock rates are the same after their
reunion, and so WGTs do not predict a SCE.

The remainder of this paper is arranged as follows. In
Section~\ref{sec:WCST}, we outline the basic properties a Weyl--Cartan
spacetime, review the existing arguments for the presence of a SCE,
and discuss an alternative approach to determining the scaling
dimension of the tangent vector to an observer's worldline. In
Section~\ref{sec:wgt}, we then discuss Weyl gauge theories of gravity,
and describe their geometric interpretation in terms of Weyl--Cartan
spacetime in Section~\ref{sec:geo}, highlighting in particular the
identification of the Weyl covariant derivative. We then reconsider
the SCE in the context of the geometric interpretation of WGTs in
Section~\ref{sec:sceinwgt}, both in terms of the norms of parallel
transported vectors and in terms of an appropriately defined proper
time. Finally, we present our conclusions in Section~\ref{sec:conc}.

%%%%%%%%%%%%%%%%%%%%%%%%%%%%%%%%%%%%%%%%%%%%%%%%%%%%%%%%%%%%%%%%%%%%%%%%%%
\section{Weyl--Cartan spacetime}
\label{sec:WCST}

\subsection{Mathematical background}

A Weyl--Cartan spacetime $Y_4$ is a differentiable manifold endowed
with a metric and affine connection, whose components in some
arbitrary holonomic coordinate system we denote by $g_{\mu\nu}$ and
${\Gamma^\mu}_{\rho\sigma}$, respectively. The latter defines a
covariant derivative operator $\nabla_\mu = \partial_\mu +
{\Gamma^\sigma}_{\rho\mu} {\matri{X}^\rho}_\sigma $, where
${\matri{X}^\rho}_\sigma$ are the $\mbox{GL}(4,R)$ generator matrices
appropriate to the tensor character under general coordinate
transformations (GCT) of the quantity to which $\nabla_\mu$ is
applied. In particular, a $Y_4$ spacetime is defined by the
requirement that this derivative operator satisfies the semi-metricity
condition
\be
\nabla_\sigma g_{\mu\nu} = -2B_\sigma g_{\mu\nu},
\label{wgtsemimet1}
\ee
where $B_\mu$ is the Weyl potential (and we have included a factor of $-2$
for later convenience and to be consistent with the notation typically
used in WGT). As Weyl originally showed, on performing the
simultaneous (gauge) transformations
\begin{subequations}
\label{eqn:weylgt}
\bea 
\bar{g}_{\mu\nu} & = & e^{2\rho} g_{\mu\nu},\\
\bar{B}_\mu & = & B_\mu - \partial_\mu\rho, 
\eea
\end{subequations}
where $\rho = \rho(x)$ is an arbitrary scalar function, the condition
(\ref{wgtsemimet1}) is preserved, i.e., one has $\nabla_\sigma
\bar{g}_{\mu\nu} = -2\bar{B}_\sigma \bar{g}_{\mu\nu}$. Thus, these
transformations define an equivalence class of Weyl--Cartan manifolds,
all of which share the same connection.  In this sense, the only
geometrical quantities of real physical significance in Weyl--Cartan
spacetime are those that transform covariantly under the
transformations (\ref{eqn:weylgt}), which may be interpreted as a
change in the length scale at every point of the manifold. It is worth
noting that if the Weyl potential is a pure gradient $B_\mu =
\partial_\mu\rho$, then the gauge transformations (\ref{eqn:weylgt})
reduce Weyl--Cartan spacetime $Y_4$ into a Riemann--Cartan $U_4$
spacetime, since $\bar{B}_\mu = 0$. More generally, if and only if the
field strength $H_{\mu\nu}=2\partial_{[\mu}B_{\nu]}$ vanishes, a $Y_4$
spacetime can be reduced to $U_4$ by a suitable transformation of the
form (\ref{eqn:weylgt}).

From (\ref{wgtsemimet1}), one finds that the connection is given
by
\begin{equation}
{\Gamma^\lambda}_{\mu\nu} = \christoffel{\ast\lambda}{\mu}{\nu}
+ {K^{\ast\lambda}}_{\mu\nu},
\label{weylaffinec}
\end{equation}
where the first term on the RHS in symmetric in $(\mu,\nu)$ and reads
\be
\christoffel{\ast\lambda}{\mu}{\nu} = \christoffel{\lambda}{\mu}{\nu}
+ \delta^\lambda_\nu B_\mu +
\delta^\lambda_\mu B_\nu - g_{\mu\nu}B^\lambda,
\label{diracconnection}
\ee
in which $\christoffel{\lambda}{\mu}{\nu}\equiv
\tfrac{1}{2}g^{\lambda\rho}(\partial_\mu g_{\nu\rho}+\partial_\nu
g_{\mu\rho}-\partial_\rho g_{\mu\nu})$ is the standard metric
(Christoffel) connection. The term ${K^{\ast\lambda}}_{\mu\nu}$ in 
(\ref{diracconnection}) is the
$Y_4$ contortion tensor, which is given in terms of (minus) the
$Y_4$ torsion ${T^{\ast\lambda}}_{\mu\nu}
= 2{\Gamma^\lambda}_{[\nu\mu]}$ by
\begin{equation}
{K^{\ast\lambda}}_{\mu\nu}=-\tfrac{1}{2}({T^{\ast\lambda}}_{\mu\nu}-
{{{T^\ast}_\nu}^{\lambda}}_\mu + {{T^\ast}_{\mu\nu}}^{\lambda}),
\label{wgtcontortiondef}
\end{equation}
and has the anti-symmetry property $K^\ast_{\lambda\mu\nu} = -
K^\ast_{\mu\lambda\nu}$ (we have placed asterisks on several
quantities above and included a minus sign in the definition of the
torsion to be consistent with the usual notation adopted in WGT). It
is clear from (\ref{weylaffinec}--\ref{wgtcontortiondef}) that setting
the torsion to zero, which is a properly invariant condition under the
gauge transformations (\ref{eqn:weylgt}), then Weyl--Cartan spacetime
$Y_4$ reduces to Weyl spacetime $W_4$.

\subsection{Physical motivation}

Aside from Weyl's original paper, much of the interest in
Weyl(--Cartan) spacetimes stems from the work of Ehlers, Pirani \&
Schild (EPS) \cite{Ehlers12}, who proposed an axiomatic approach to
determining a suitable mathematical model of spacetime using only
basic assumptions about the behavior of freely falling massive and
massless particles. This led to the conclusion that massless particles
determine a conformal structure on spacetime, while the massive
particles determine a projective structure on spacetime. By imposing a
compatibility condition on these two structures, basically postulating
that massless particle trajectories can be approximated arbitrarily
well by massive particle ones, EPS arrived at Weyl spacetime as the
appropriate mathematical model. In their approach, EPS assumed the
connection to be symmetric (torsionless) from the outset, but their
conclusions rely only on the semi-metricity condition
(\ref{wgtsemimet1}) and so it is reasonable to consider the more
general Weyl--Cartan spacetime, which allows for non-zero torsion.

\subsection{Parallel transported vectors and the SCE}

Using the semi-metricity condition (\ref{wgtsemimet1}), which holds
irrespective of the presence of torsion, the evolution of the inner
product of any two vectors $v^\mu$ and $w^\mu$ parallel transported
along some curve $\mathcal{C}$ is given by
\bea
\nd{}{\lambda}[g_{\mu\nu}v^\mu(\lambda)w^\nu(\lambda)] 
&=& \frac{D}{D\lambda}[g_{\mu\nu}v^\mu(\lambda)w^\nu(\lambda)],\nonumber\\
&=& (u^\sigma(\lambda)\nabla_\sigma g_{\mu\nu}) v^\mu(\lambda) w^\nu(\lambda)  \nonumber\\
&=& -2B_\sigma u^\sigma(\lambda) g_{\mu\nu} v^\mu(\lambda) w^\nu(\lambda),
\label{eq:normevol}
\eea
where $u^\mu(\lambda) = dx^\mu/d\lambda$ is the tangent vector to
${\cal C}$ at the parameter value $\lambda$ and we have used the
parallel transport conditions $Dv^\mu/D\lambda = 0 =
Dw^\mu/D\lambda$. Hence, on integrating, the inner product as a
function of the parameter $\lambda$ along the curve is given by
\be
g_{\mu\nu}v^\mu(\lambda)w^\nu(\lambda) = 
g_{\mu\nu}v^\mu(\lambda_0)w^\nu(\lambda_0)e^{-2\int_{\lambda_0}^\lambda
B_\sigma u^\sigma(\lambda') d\lambda'}.
\label{eqn:innerprod}
\ee
By setting $v^\mu=w^\mu$ and considering parallel transport around a
closed curve ${\cal C}$, the length $\ell$ of a vector on completing a
loop is related to its original length $\ell_0$ by
\be \ell = \ell_0\,e^{-\oint_{\cal C} B_\mu dx^\mu}.  
\ee
Thus, using Stokes' theorem, the condition $\ell = \ell_0$ holds if
and only if $H_{\mu\nu}$ vanishes throughout the region interior to
${\cal C}$.  The above result forms the basis of the original
discussions of the SCE in Weyl's theory by Einstein
and others, as described in the Introduction.

\subsection{Proper time and the SCE}
\label{sec:propertime}

As also mentioned in the Introduction, however, the intuitive argument
above is not rigorous, and a more careful approach is based on
defining a consistent notion of proper time along (timelike)
worldlines in Weyl spacetimes \cite{Ehlers12, Perlick87, Avalos18}.
The simplest construction is based on the requirement that for a
timelike curve $x^\mu(\tau)$ to be parameterised by proper time
$\tau$, one requires the tangent vector $u^\mu(\tau) = dx^\mu/d\tau$,
which is thus the particle 4-velocity, to be orthogonal to its
4-acceleration $a^\mu(\tau) = Du^\mu/D\tau$, such that
\be
g_{\mu\nu}u^\mu(\tau)a^\nu(\tau) = 0.
\label{eqn:clockdef}
\ee

This derivation of the proper time $\tau$ is unaffected by the
presence of non-zero torsion, since it also depends only on the
semi-metricity condition (\ref{wgtsemimet1}), and is hence applicable
in Weyl--Cartan spacetimes. It is useful first to note
that, for two {\it arbitrary} parameterisations $\lambda$ and $\xi$ of
a worldline, the corresponding tangent vectors and their absolute
derivatives (which should no longer strictly be interpreted as the
particle 4-velocity and acceleration) are related by \cite{Avalos18}
\begin{subequations}
\bea
u^\mu(\lambda) &=& \nd{\xi}{\lambda}u^\mu(\xi),\\
a^\mu(\lambda) &=& \hnd{\xi}{\lambda}{2}u^\mu(\xi)
+ \left(\nd{\xi}{\lambda}\right)^2 a^\mu(\xi).
\eea
\end{subequations}
By considering the quantity $g_{\mu\nu}u^\mu(\lambda)a^\nu(\lambda)$
and making the identification $\xi=\tau$, for which we require
(\ref{eqn:clockdef}) to hold, one finds
\be
\hnd{\tau}{\lambda}{2} 
- \frac{g_{\mu\nu}u^\mu(\lambda)a^\nu(\lambda)}{g_{\mu\nu}u^\mu(\lambda)u^\nu(\lambda)}
\nd{\tau}{\lambda}=0.
\label{eqn:tauode}
\ee
Since this a linear differential equation, if $\tau$ is a solution,
then so too is $a\tau + b$, where $a$ and $b$ are
constants that represent merely the scaling and zero point of the proper
time variable, respectively. To proceed further, it is convenient to
consider the quantity
\bea
\nd{}{\lambda}[g_{\mu\nu}u^\mu(\lambda)u^\nu(\lambda)] 
&=& \frac{D}{D\lambda}[g_{\mu\nu}u^\mu(\lambda)u^\nu(\lambda)],\nonumber\\
&=& (u^\sigma\nabla_\sigma g_{\mu\nu}) u^\mu u^\nu  + 2g_{\mu\nu}u^\mu
a^\nu,\nonumber\\
&=& -2B_\sigma u^\sigma g_{\mu\nu} u^\mu u^\nu  + 2g_{\mu\nu}u^\mu
a^\nu,\phantom{A\nd{}{\lambda}}
\label{eqn:taucalc1}
\eea
where we use the semi-metricity condition (\ref{wgtsemimet1}), and in
the last two lines (and the remainder of this section) we drop the
explicit dependence of quantities on the arbitrary parameter $\lambda$
for brevity.
%and in the final equality we have used the
%semi-metricity condition (\ref{wgtsemimet1}), which holds irrespective
%of the presence of torsion. 
Thus, (\ref{eqn:tauode}) becomes
\be \hnd{\tau}{\lambda}{2} -
\left[\tfrac{1}{2}\nd{}{\lambda}(g_{\mu\nu}u^\mu u^\nu)+g_{\mu\nu}B^\mu
u^\nu\right] \nd{\tau}{\lambda}=0,
\label{eqn:tauode2}
\ee
which is straightforwardly solved to obtain the proper time interval
$\Delta\tau$ between two events corresponding to the parameter values
$\lambda_0$ and $\lambda$ along the worldline:
\be
\Delta\tau(\lambda) =
\left.\frac{d\tau/d\lambda}{\sqrt{g_{\mu\nu}u^\mu
    u^\nu}}\right|_{\lambda_0} 
\int_{\lambda_0}^\lambda e^{\int_{\lambda_0}^{\lambda'} B_\mu u^\mu d\lambda''} \sqrt{g_{\mu\nu}u^\mu
    u^\nu}\,d\lambda'.
\label{eqn:deltataudef}
\ee

The application of this result to the two-clock thought experiment
discussed in the Introduction is straightforward. Suppose the two
clocks are separated at some event and thereafter follow the
worldlines ${\cal C}_1$ and ${\cal C}_2$, respectively, before being
reunited at some other event. The ratio of the elapsed proper time
measured by each clock between their reunion and some subsequent event
along their joint worldline is given by
\be
\frac{\Delta\tau_2}{\Delta\tau_1} =  \exp\left(\int_{{\cal C}_2} B_\mu dx^\mu
  - \int_{{\cal C}_1} B_\mu dx^\mu\right).
\ee
Thus, in general, the clock rates differ after their reunion and so
Weyl--Cartan spacetime exhibits a SCE.

\subsection{Weyl weight of worldline tangent vector}
\label{sec:uweight}

The above derivation of the proper time $\tau$ is based on the
condition (\ref{eqn:clockdef}), which is assumed to be consistent across
the whole equivalence class defined by the Weyl gauge transformations
(\ref{eqn:weylgt}). This consistency holds, however, only if the
quantities $u^\mu(\tau)$ transform covariantly with Weyl weight $w=0$
(i.e. they are invariant) under these transformations. \myNew{As discussed in the Introduction,
  however, we argue that these quantities in fact have weight $w=-1$,
  based on consideration of their corresponding components in the
  tetrad basis. One may, however, obtain further insight into this
  conclusion without the use of tetrads, as outlined below.}

One may in fact work more generally in terms of an arbitrary parameter
$\lambda$, such that $u^\mu(\lambda) = dx^\mu/d\lambda$ is the tangent
vector at the parameter value $\lambda$ to the worldline $\mathcal{C}$
given by $x^\mu=x^\mu(\lambda)$. Since the coordinates are
unaffected by the gauge transformations (\ref{eqn:weylgt}), they have
weights $w = 0$, so it remains only to determine the weight of
the operator $d/d\lambda$. One may achieve this by first writing
\be
\nd{}{\lambda} = \nd{s}{\lambda}\nd{}{s},
\ee
where $ds^2 = g_{\mu\nu}\,dx^\mu\,dx^\nu$. Similarly, after the gauge
transformations (\ref{eqn:weylgt}), one has
\be
\myNew
{\frac{d}{d\bar{\lambda}} = \nd{\bar{s}}{\bar{\lambda}}\nd{}{\bar{s}}},
\ee
By requiring that \myNew{$d\bar{s}/d\bar{\lambda} = ds/d\lambda$}, so that the
\myNew{physical} distance\myNew{, as opposed to the coordinate
  distance,} along the curve ${\cal C}$ is traced out at the same rate
before and after the gauge transformations (\ref{eqn:weylgt}), and
using the fact that $d\bar{s} = e^\rho\,ds$, then
\be
\frac{d}{d\bar{\lambda}}
= e^{-\rho}\nd{s}{\lambda}\nd{}{s} = e^{-\rho}\nd{}{\lambda}.
\ee
Thus, the tangent vector $u^\mu(\lambda) = dx^\mu/d\lambda$ has Weyl
weight $w=-1$. \myNew{It is worth noting that this conclusion does not
  mean that the worldline $\mathcal{C}$ changes under the gauge
  transformation, but only that the coordinates along it are traced
  out at a different rate before and after the transformation. For
  example, if two points $O$ and $A$ on the curve with coordinates
  $x^\mu_0$ and $x^\mu_A$ correspond to parameter
  values $\lambda = 0$ and $\lambda = \lambda_A$, respectively, before
  the gauge transformation, then these points will have the same
  coordinates but parameter
  values $\bar{\lambda} = 0$ and $\bar{\lambda} \neq \lambda_A$ after
  the transformation. The most important consequence of
  $u^\mu(\lambda)$ having weight $w=-1$ is, however, that} the length
of the tangent vector is invariant under the gauge transformations
(\ref{eqn:weylgt}), which follows immediately since
\be
g_{\mu\nu} u^\mu u^\nu \to \bar{g}_{\mu\nu} \bar{u}^\mu \bar{u}^\nu 
= g_{\mu\nu} u^\mu u^\nu,
\ee
%
%the results of applying the Weyl covariant derivative
%$\nabla^\ast_\sigma$ and the standard
%$Y_4$ covariant derivative $\nabla_\sigma$ to the components $u^\mu(\lambda)$ will
%differ.
whereas the condition (\ref{eqn:clockdef}) (but with $\tau$ replaced
by $\lambda$) is not invariant, since one may show that
\be
\bar{g}_{\mu\nu}\bar{u}^\mu\bar{a}^\nu 
= e^{-\rho} g_{\mu\nu}u^\mu[a^\nu - u^\sigma(\partial_\sigma\rho)u^\nu].
\ee
%

%
\begin{comment}
Alternative derivation:
\bea
\bar{u}^\mu(\bar{\lambda}) 
& = &  \nd{x^\mu}{\bar{s}}\nd{\bar{s}}{\bar{\lambda}} \\
& = &  \nd{x^\mu}{\bar{s}}\nd{s}{\lambda} \\
& = &  e^{-\rho}\nd{x^\mu}{s}\nd{s}{\lambda} \\
& = &  e^{-\rho}u^\mu(\lambda)
\eea
\end{comment}
%

This lack of invariance of the condition (\ref{eqn:clockdef}) under
Weyl gauge transformations undermines the physical significance of the
proper time variable derived above. As we show below, however, by
considering the geometric interpretation of Weyl gauge theories of
gravity, one may identify an alternative form of covariant derivative
which both leaves the length of a vector unchanged after parallel
transport around a closed loop and allows one to define an analogous
condition to (\ref{eqn:clockdef}) which is invariant under Weyl gauge
transformations.

%%%%%%%%%%%%%%%%%%%%%%%%%%%%%%%%%%%%%%%%%%%%%%%%%%%%%%%%%%%%%%%%%%%%%%%%%%
\section{Weyl gauge theories of gravity}
\label{sec:wgt}

It was the gauging of the Poincar\'e group $\mathcal{P}$ by
Kibble \cite{Kibble61} that first revealed how to achieve a meaningful
gauging of groups that act on the points of spacetime as well as on
the components of physical fields. The essense of Kibble's approach
was to note that when the parameters of the Poincar\'e group become
independent arbitrary functions of position, this leads to a complete
decoupling of the translational parts from the rest of the group, and
the former are then interpreted as arising from a general coordinate
transformation (GCT; or spacetime diffeomorphisms, if interpreted
actively). Thus the action of the gauged Poincar\'e group is
considered as a GCT $x^\mu \to x^{\prime\mu}$, together with the local
action of its Lorentz subgroup on the orthonormal tetrad
basis vectors $\hat{\vect{e}}_a(x)$ that define local Lorentz
reference frames, where we adopt the common convention that Latin
indices (from the start of the alphabet) refer to anholonomic local
tetrad frames, while Greek indices refer to holonomic coordinate
frames.  This approach leads to Poincar\'e gauge theories (PGT) of
gravity, but can be straightforwardly extended to more general
spacetime symmetry groups \cite{Harnad76,Lord86a,Blagojevic02}.

A natural extension of PGT is also to demand local scale invariance,
which is most directly achieved by gauging the Weyl group
$\mathcal{W}$ \cite{Bregman73,Charap74,Kasuya75}. This may be
formulated in a number of ways, e.g. by considering the Weyl
transformations as active or passive, infinitesmial or finite, but
they are all essentially equivalent. As in PGT, the physical model is
an underlying Minkowski spacetime in which a set of matter fields
$\vpsi_i$ is distributed continuously (these fields may include a
scalar compensator field, which we occasionally also denote by
$\phi$). Since the spacetime is Minkowski, one may adopt a global
Cartesian inertial coordinate system $x^\mu$, which greatly simplifies
calculations, but more general coordinate systems may be
straightforwardly accommodated, if required \cite{eWGTpaper}.
The field dynamics are described by a matter action 
\be
S_{\rm M} = \int L_{\rm M}(\vpsi_i,\partial_\mu \vpsi_i)\,d^4x, 
\ee
which is invariant under the global action of the Weyl group.  One
then gauges the Weyl group $\mathcal{W}$ by demanding that the matter
action be invariant with respect to (infinitesimal, passively
interpreted) GCT and the local action of the subgroup $\mathcal{H}$
(the homogeneous Weyl group), obtained by setting the translation
parameters of $\mathcal{W}$ to zero (which leaves the origin $x^\mu=0$
invariant), and allowing the remaining group parameters to become
independent arbitrary functions of position.

In this way, one is led to the introduction of new field variables
${h_a}^{\mu}$, ${A^{ab}}_{\mu}$ and $B_\mu$, corresponding to the
translational, rotational and dilational parts of the Weyl group,
respectively. These fields are interpreted as gravitational gauge
fields and are used to assemble the covariant derivative (adopting the
common notation in WGT \cite{Blagojevic02,eWGTpaper})
\be
{\cal D}^\ast_a\vpsi_i = {h_a}^\mu D^\ast_\mu \vpsi_i =
{h_a}^\mu (\partial_\mu +
\tfrac{1}{2}{A^{ab}}_\mu\Sigma_{ab}+w_iB_\mu)\vpsi_i,
\ee
where the field $\vpsi_i$ is assumed to have Weyl weight $w_i$ and
$\Sigma_{ab} = -\Sigma_{ba}$ are the generators matrices of the
$\mbox{SL}(2,C)$ representation to which $\vpsi_i$ belongs. Since
${\cal D}^\ast_a\vpsi_i$ is constructed to transform in the same way
under the action of the gauged group $\mathcal{W}$ as
$\partial_\mu\vpsi_i$ does under the global action of $\mathcal{W}$,
the matter action in the presence of gravity is then typically
obtained by the minimal coupling procedure of replacing partial
derivatives in the special-relativistic matter Lagrangian by covariant
ones, to obtain 
\be
S_{\rm M} = \int h^{-1} L_{\rm M}(\vpsi_i,{\cal
  D}_a\vpsi_i)\,d^4x,
\ee
where the factor containing $h \equiv \mbox{det}({h_a}^\mu)$ is
required to make the integrand a scalar density rather than a
scalar. It should noted that the requirement of local scale invariance
imposes tight constraints on the precise form of $L_{\rm M}$. In
particular, since $h^{-1}$ has a Weyl (or conformal) weight
$w(h^{-1})=4$, the Lagrangian $L_{\rm M}$ must have a weight $w(L_{\rm
  M})=-4$.

In addition to the matter action, the total action must also contain
terms describing the dynamics of the free gravitational gauge fields.
The latter are constructed from the field strength tensors ${\cal
  R}_{abcd}$, ${\cal T}^\ast_{abc}$, ${\cal H}_{ab}$ of the rotational,
translational and dilational gauge fields, respectively,
which are defined through the action of the commutator
of two covariant derivatives on some field $\vpsi$ of weight $w$ by
\be
[{\cal D}^\ast_c,{\cal D}^\ast_d]\vpsi =
\tfrac{1}{2}{{\cal R}^{ab}}_{cd}\Sigma_{ab}\vpsi + w{\cal H}_{cd}\vpsi
- {{\cal T^\ast}^a}_{cd}{\cal D}^\ast_a\vpsi.
\ee
It is straightforward to show that the fields strengths have the forms
${{\cal R}^{ab}}_{cd} = {h_a}^{\mu}{h_b}^{\nu}{R^{ab}}_{\mu\nu}$,
${\cal H}_{cd} = {h_c}^\mu {h_d}^\nu H_{\mu\nu}$ and ${{\cal T}^{\ast
    a}}_{bc} = {h_b}^{\mu}{h_c}^{\nu} {T^{\ast a}}_{\mu\nu}$, where
\bea
{R^{ab}}_{\mu\nu}  & = & 
 2(\partial_{[\mu} {A^{ab}}_{\nu]} +
\eta_{cd}{A^{ac}}_{[\mu}{A^{db}}_{\nu]}),
%\partial_\mu {A^{ab}}_\nu \!-\! \partial_\nu {A^{ab}}_\mu
%\!+\!{A^a}_{c\mu}{A^{cb}}_\nu \!-\! {A^a}_{c\nu}{A^{cb}}_\mu,
%\phantom{AA}
\label{rfsdef} \\
H_{\mu\nu} & = & 2\partial_{[\mu} B_{\nu]},
\label{dfsdef}\\
{T^{\ast a}}_{\mu\nu} & = & 2D^\ast_{[\mu} {b^a}_{\nu]},
\label{tfsdef}
\eea
and ${b^a}_\mu$ is the inverse of ${h_a}^\mu$. It is worth noting that 
${{\cal R}^{ab}}_{cd}$ has the same functional form as the rotational
field strength in PGT (which we thus denote by the same symbol), but
that ${{\cal T}^{\ast a}}_{bc} =  {{\cal T}^a}_{bc} + \delta^a_c{\cal
  B}_b - \delta^a_b{\cal B}_c$, where ${{\cal T}^a}_{bc}$ is the
translational field strength in PGT and ${\cal B}_a = {h_a}^\mu B_\mu$. 
The free gravitational action then has the
general form 
\be S_{\rm G} = \int h^{-1} L_{\rm G}({{\cal R}^{ab}}_{cd},{{\cal
    T}^{\ast a}}_{bc},{\cal H}_{ab})\,d^4x, 
\ee 
where $L_{\rm G}$ must also have a Weyl (conformal) weight $w(L_{\rm
  G})=-4$, which places tight constraints on its form.  It is easily
shown that $w({{\cal R}^{ab}}_{cd})=w({\cal H}_{ab})=-2$ and $w({{\cal
    T}^{\ast a}}_{bc})=-1$, which means that $L_{\rm G}$ can be
quadratic in ${{\cal R}^{ab}}_{cd}$ and ${\cal H}_{ab}$, while terms
linear in ${\cal R} \equiv {{\cal R}^{ab}}_{ab}$ or quadratic in
${{\cal T}^{\ast a}}_{bc}$ are not allowed, despite them transforming
covariantly under local Weyl transformations. 
%Thus, in WGT, the
%general form of $L_{\rm G}$, possessing terms no higher than quadratic
%order in the field strengths, and hence at most quadratic in the first
%derivatives of gauge fields, is of ${\cal R}^2+{\cal H}^2$ type:
%%
%\begin{widetext}
%\begin{equation}
%L_{\rm G}   = \alpha_1 {\cal R}^2
%+ \alpha_2 {\cal R}_{ab}{\cal R}^{ab}
%+ \alpha_3 {\cal R}_{ab}{\cal  R}^{ba}
%+ \alpha_4 {\cal R}_{abcd}{\cal R}^{abcd}
%+ \alpha_5 {\cal R}_{abcd}{\cal R}^{acbd}
%+ \alpha_6 {\cal R}_{abcd}{\cal R}^{cdab}
%+ \xi {\cal H}_{ab}{\cal H}^{ab} \equiv  L_{{\cal R}^2} + L_{{\cal H}^2},
%\label{lg2}
%\end{equation}
%\end{widetext}
%%
%where the $\alpha_i$ and $\xi$ are dimensionless free parameters. Once
%again pseudo-scalar terms have been omitted by requiring parity
%invariance, and one may use the generalised Gauss--Bonnet identity to
%set one of $\alpha_1$, $\alpha_3$ or $\alpha_6$ to zero, without loss
%of generality.
%Evidently, local Weyl
%invariance removes many of the possibilities that exist in PGT;
%essentially the ${\cal R}$ and ${\cal T}^{\ast 2}$ terms, which
%possess dimensional constants, are forbidden.
%Although the requirement that each term in the total Lagrangian must
%have a Weyl weight $w =-4$ clearly places quite restrictive conditions
%on its form, 

One can, however, construct further Weyl-covariant terms with the
appropriate weight for inclusion in the total Lagrangian by
introducing an additional massless scalar field (or fields) $\phi$
with Weyl weight $w({\phi})=-1$, often termed the compensator(s) \cite{Blagojevic02}. This opens up possibilities for the inclusion of
further action terms in which the scalar field is non-minimally
(conformally) coupled to the field strength tensors of the
gravitational gauge fields, combined (usually) with an additional free
kinetic term for $\phi$. For example, terms proportional to $\phi^2
{\cal R}$ or $\phi^2 L_{{\cal T}^{\ast 2}}$, where $L_{{\cal T}^{\ast
    2}}$ consists of terms quadratic in ${{\cal T}^{\ast a}}_{bc}$,
are Weyl-covariant with weight $w=-4$ and so may be added to the total
Lagrangian \cite{Dirac73,Omote77,Sijacki82,Neeman88}.
%In any case, the
%resulting total Lagrangian is at most quadratic in the first
%derivatives of the gauge fields, thereby satisfying the
%pseudolinearity hypothesis that field equations be linear in the
%second-order derivatives of the gauge field, and hence ensuring that
%such theories do not suffer from Ostrogradsky's instability.

The inclusion of scalar fields also allows for more flexibility in the
allowed forms of the matter. An important example is a free Dirac
field $\psi$, which has Weyl weight
$w(\psi)=w(\bar{\psi})=-\tfrac{3}{2}$, and for which the Lagrangian
reads
\be L_{\rm D} =
\tfrac{1}{2}i\bar{\psi}\gamma^\mu{\stackrel{\leftrightarrow}{\partial_\mu}}\psi
- m\bar{\psi}\psi.
\label{diraclm2}
\ee
The corresponding action is not scale-invariant owing to the mass term
$m\bar{\psi}\psi$.  It thus appears that one requires the field to be
massless, which clearly cannot describe `ordinary' matter.  This
difficulty can be circumvented, however, by making the replacement
$m\bar{\psi}\psi \to \mu\phi\bar{\psi}\psi$, where $\mu$ is a
dimensionless parameter but $\mu\phi$ has the dimensions of mass in
natural units. The action is then invariant under the global Weyl
group, and one may also add kinetic and self-interaction terms of
weight $w=-4$ for the scalar field $\phi$. After gauging the Weyl
group as outlined above, the resulting WGT matter Lagrangian
for the Dirac and compensator scalar field is given by
\be
L_{\rm M} = \tfrac{1}{2}i\bar{\psi}\gamma^a
{\stackrel{\leftrightarrow}{{\cal D}_a}}\psi - \mu\phi\bar{\psi}\psi
+ \tfrac{1}{2}\nu ({\cal D}^\ast_a\phi) ({\cal D}^{\ast a} \phi) -
\lambda\phi^4,
\label{weylcovardirac3}
\ee
where $\mu$, $\nu$ and $\lambda$ are dimensionless constants (usually
positive). In this way, although the trace of the total energy
momentum tensor of the $\psi$ and $\phi$ fields must vanish, the
energy-momentum tensor of the Dirac matter field $\psi$ itself need
not be traceless, thereby allowing it to be massive. Indeed, this
approach to the construction of gauge theories of gravity that are
scale-invariant but, at the same time, are able to accommodate
`ordinary' matter was first explored by Dirac \cite{Dirac73}.

More generally, the introduction of scalar fields in WGT is
also important since they provide a natural means for spontaneously
breaking the scale symmetry. The approach most commonly adopted is to
use local scale invariance to set the compensator scalar field $\phi$
to a constant value in the resulting field equations, which is known
as the Einstein gauge. Setting $\phi=\phi_0$ in the equation of motion
for the Dirac field $\psi$, for example, leads to its interpretation
as a massive field with $m=\mu\phi_0$. It is usually considered that
setting $\phi=\phi_0$ represents the choice of some definite scale in
the theory, thereby breaking scale-invariance. Indeed, it is often
given the physical interpretation of corresponding to some spontaneous
breaking of the scale symmetry (where Nature chooses the gauge).  This
interpretation is questionable, however, since the equations of motion
in the Einstein gauge are identical in form to those obtained when
working in scale-invariant variables, where the latter involves no
breaking of the scale symmetry \cite{eWGTpaper}.

In any case, the total action is taken as the sum of the matter and
gravitational actions, and variation of the total action with respect
to the gauge fields ${h_a}^{\mu}$, ${A^{ab}}_{\mu}$ and $B_\mu$ leads
to three coupled gravitational field equations in which the
energy-momentum ${\tau^k}_\mu\equiv \delta{\cal L}_{\rm M}/\delta
{h_a}^\mu$, spin-angular-momentum ${\sigma_{ab}}^\mu\equiv \delta{\cal
  L}_{\rm M}/\delta {A^{ab}}_\mu$ and dilation current $\zeta^\mu
\equiv \delta {\cal L}_{\rm M}/\delta B_\mu$ of the matter field act
as sources, where ${\cal L}_{\rm M} \equiv h^{-1} L_M$.

%%%%%%%%%%%%%%%%%%%%%%%%%%%%%%%%%%%%%%%%%%%%%%%%%%%%%%%%%%%%%%%%%%%%%%%%%%
\section{Geometric interpretation of WGT}
\label{sec:geo}

Kibble's gauge approach to gravitation is most naturally interpreted
as a field theory in Minkowski
spacetime \cite{Wiesendanger96,Lasenby98,eWGTpaper}, in the same way as
the gauge field theories describing the other fundamental
interactions.  It is more common, however, to reinterpret the
mathematical structure of gravitational gauge theories
geometrically \cite{Blagojevic02}.

At the heart of the geometric interpretation is the identification of
${h_a}^\mu$ as the components of a vierbein system in a more general
spacetime. Thus, at any point $x$ in the spacetime, one demands that
the orthonormal tetrad frame vectors $\hvect{e}_a(x)$ and the
coordinate frame vectors $\vect{e}_\mu(x)$ are related
by
\begin{equation}
\hvect{e}_a = {h_a}^\mu\vect{e}_\mu,\qquad \vect{e}_\mu =
{b^a}_\mu\hvect{e}_a,
\label{geotetrad}
\end{equation}
with similar relationships holding between the dual basis vectors
$\hvect{e}^a(x)$ and $\vect{e}^\mu(x)$ in each set.  For any other
vector $\vect{V}$, written in the coordinate basis as (say)
$V_\mu\vect{e}^\mu$, one then identifies the quantities $V_a =
{h_a}^\mu V_\mu$, for example, as the components of the {\it same}
vector, but in the tetrad basis.  This is a fundamental difference
from the Minkowski spacetime viewpoint presented earlier, in which
${\cal V}_a={h_a}^\mu V_\mu$ would be regarded as the components in
the tetrad basis of a {\it new} vector field ${\cal V}$.

The identification of ${h_a}^\mu$ as the components of a vierbein
system has a number of far-reaching consequences. Firstly, the
index-conversion properties of ${h_a}^\mu$ and ${b^a}_\mu$ are
extended. It is straightforward to show, for example, that ${h_a}^\mu
V^a=V^\mu$ and ${b^a}_\mu V^\mu=V^a$. Moreover, any contraction over
Latin (Greek) indices can be replaced by one over Greek (Latin)
indices. None of these operations is admissible when the $h$, $A$ and
$B$ fields are viewed purely as gauge fields in Minkowski spacetime.

Perhaps the most important consequence of identifying ${h_a}^\mu$ as
the components of a vierbein system is that the inner product of the
coordinate basis vectors becomes
\begin{equation}
\vect{e}_\mu \cdot \vect{e}_\nu = \eta_{ab}{b^a}_\mu {b^b}_\nu \equiv g_{\mu\nu}.
\label{metricdef}
\end{equation}
Thus, in this geometric interpretation, one must work in a more
general spacetime with metric $g_{\mu\nu}$.
Conversely, since the tetrad basis vectors still form an orthonormal set,
one has
\begin{equation}
\hvect{e}_a \cdot \hvect{e}_b = \eta_{ab} = g_{\mu\nu}{h_a}^\mu {h_b}^\nu.
\label{etagdef}
\end{equation}
From (\ref{metricdef}), one also finds that $h^{-1}=\sqrt{-g}$ (where
we are working with a metric signature of $-2$).  Under a (local,
physical) dilation, the spacetime metric and $h$-field have Weyl
weights $w(g_{\mu\nu})=2$ and $w({h_a}^\mu)=-1$ respectively, and so
(\ref{metricdef}) and (\ref{etagdef}) imply that $w(\eta_{ab})=0$, as
expected. From (\ref{geotetrad})--(\ref{etagdef}), one immediately
finds that the $h$-field and its inverse are directly related by index
raising/lowering, so there no need to distinguish between them by
using different kernel letters. Consequently, the standard practice,
which we will follow here, is to notate ${h_a}^\mu$ and ${b^a}_\mu$ as
${e_a}^\mu$ and ${e^a}_\mu$, respectively.  One should also note that,
if the components $V_\mu$ and $V^\mu$ have Weyl weights $w$ and
$\tilde{w} = w-2$, respectively, then the components $V_a$ and $V^a$
have weights $w-1$ and $\tilde{w}+1 = w-1$.

One is also led naturally to the interpretation of ${A^{ab}}_\mu$ as
the components of the `spin-connection' that encodes the rotation
of the local tetrad frame between points $x$ and $x+\delta x$, which
is accompanied by a local change in the standard of length between the
two points, which is encoded by $B_\mu$. Thus, the operation of
parallel transport for some vector $V^a$ of weight $w$ is defined as
\begin{equation}
\delta V^a = - ({A^a}_{b\mu}+wB_\mu\delta^a_b) V^b \,\delta x^\mu,
\label{weylparallel}
\end{equation}
which is required to compare vectors
$V^a(x)$ and $V^a(x+\delta x)$ at points $x$ and $x+\delta x$, determined with
respect to the tetrad frames $\hvect{e}_a(x)$ and
$\hvect{e}_a(x+\delta x)$ respectively.
Hence, in general, a vector not only changes its direction on parallel
transport around a closed loop, but also its length. The expression
(\ref{weylparallel}) establishes the correct form for the related
$(\Lambda,\rho)$-covariant derivative, e.g.
\begin{equation}
D^\ast_\mu V^a = \partial_\mu V^a + wB_\mu V^a +  {A^a}_{b\mu}V^b =
\partial^\ast_\mu V^a +  {A^a}_{b\mu}V^b,
\end{equation}
where $\partial^\ast_\mu \equiv \partial_\mu + wB_\mu$.
Moreover, the existence of tetrad frames
at each point of the spacetime implies the existence of the Lorentz
metric $\eta_{ab}$ at each point. Then demanding that $\eta_{ab}$ is
invariant under parallel transport, and recalling that
$w(\eta_{ab})=0$, requires the spin-connection to be antisymmetric,
i.e.  ${A^{ab}}_\mu=-{A^{ba}}_\mu$, as previously.

Further differences between the Minkowski spacetime gauge field
viewpoint and the geometric interpretation occur when generalising the
$(\Lambda,\rho)$-covariant derivative to apply to fields with definite
GCT tensor behaviour. First, in the geometric interpretation, one can
in general no longer construct a global inertial Cartesian coordinate
system in the more general spacetime. Thus, one must rely on arbitrary
coordinates and so define the `total' covariant derivative
\begin{equation}
\Delta^\ast_\mu \equiv \partial^\ast_\mu + {\Gamma^\sigma}_{\rho\mu} {\matri{X}^\rho}_\sigma
+ \tfrac{1}{2}{A^{ab}}_\mu\Sigma_{ab} = \nabla^\ast_\mu + D^\ast_\mu - \partial^\ast_\mu,
\label{weylextcovd2}
\end{equation}
where $\nabla^\ast_\mu = \partial^\ast_\mu + {\Gamma^\sigma}_{\rho\mu}
{\matri{X}^\rho}_\sigma = \nabla_\mu + w B_\mu$, in which  
${\matri{X}^\rho}_\sigma$ are the $\mbox{GL}(4,R)$ generator
matrices appropriate to the GCT tensor character of the field to which
$\Delta^\ast_\mu$ is applied and $w$ is the Weyl weight of the field.
If a field $\psi$ carries only Latin indices, then
$\nabla^\ast_\mu\psi =\partial^\ast_\mu\psi$ and so
$\Delta^\ast_\mu\psi=D^\ast_\mu\psi$; conversely, if a field $\psi$
carries only Greek indices, then $D^\ast_\mu\psi
=\partial^\ast_\mu\psi$ and so
$\Delta^\ast_\mu\psi=\nabla^\ast_\mu\psi$. When acting on an object of
weight $w$, for all these derivative operators the resulting object
also transforms covariantly with the same weight $w$.

Most importantly, in a dynamical spacetime, the affine connection
coefficients ${\Gamma^\sigma}_{\rho\mu}$ are themselves dynamical
variables, no longer fixed by our choice of coordinate system.  They
are, however, necessarily related to the spin-connection and dilation
vector since the tetrad components $V^a$ of a vector with coordinate
components $V^\mu$ should, when parallel transported from $x$ to
$x+\delta x$, be equal to $V^a+\delta V^a$, i.e.
\begin{equation}
V^a+\delta V^a = (V^\mu+\delta V^\mu)\,{e^a}_\mu(x+\delta x).
\label{vreln}
\end{equation}
If the vector components $V^\mu$ have Weyl weight $w$, we substitute
for $\delta V^a$ using (\ref{weylparallel}), but with $w \to w-1$, and
denote parallel transport of the coordinate basis components in an
analogous fashion by defining
\be
\delta V^\mu = -({\Gamma^\mu}_{\rho\sigma}
+ wB_\sigma\delta^\mu_\rho) V^\rho\delta x^\sigma.
\label{eq:ptdef}
\ee
In other words, the quantities ${\Gamma^\sigma}_{\rho\mu}$ contain the
same geometrical information as ${e^a}_\mu$ and ${A^{ab}}_\mu$, but in
a different frame. This information corresponds to 40 independent
gravitational field variables, and there are a further 4 variables contained in
$B_\mu$.

From (\ref{vreln}), we obtain the relation
\begin{equation}
\Delta^\ast_\mu {e^a}_\nu  \equiv
\partial^\ast_\mu {e^a}_\nu - {\Gamma^\sigma}_{\nu\mu}{e^a}_\sigma
+ {A^a}_{b\mu}{e^a}_\nu =0,
\label{weyltetradp}
\end{equation}
which relates $A$ and $\Gamma$ (and $B$); in particular, we note that
$w({\Gamma^\sigma}_{\nu\mu})=0$. The relation (\ref{weyltetradp}) is
sometimes known as the `tetrad postulate', but note that it always
holds. It is straightforward to show that $A$ or $\Gamma$ may be
written explicitly in terms of the other as
\begin{eqnarray}
{\Gamma^\lambda}_{\nu\mu} & = &
{e_a}^\lambda (\partial^\ast_\mu {e^a}_\nu +
{A^a}_{b\mu} {e^b}_\nu),\label{gammaofa}\\
{A^a}_{b\mu} & = &
{e^a}_\lambda (\partial^\ast_\mu {e_b}^\lambda
+ {\Gamma^\lambda}_{\nu\mu}{e_b}^\nu).
\label{aofgamma}
\end{eqnarray}

Using (\ref{metricdef}) and (\ref{weyltetradp}), one
finds that 
\be
\nabla^\ast_\sigma g_{\mu\nu}=0,
\label{eqn:nablastar}
\ee
and so this derivative
operator commutes with raising and lowering of coordinate
indices. Equivalently, one may write this semi-metricity condition as
\begin{equation}
\nabla_\sigma g_{\mu\nu} = -2B_\sigma g_{\mu\nu},
\label{wgtsemimet2}
\end{equation}
which shows that the spacetime has, in general, a Weyl--Cartan $Y_4$
geometry. Hence, as discussed in Section~\ref{sec:WCST}, the
connection ${\Gamma^\lambda}_{\mu\nu}$ must satisfy the conditions
(\ref{weylaffinec})--(\ref{wgtcontortiondef}).

Moreover, substituting (\ref{aofgamma}) into the
expressions (\ref{rfsdef})--(\ref{tfsdef}) for the gauge field
strengths ${R^{ab}}_{\mu\nu}$, $H_{\mu\nu}$ and ${T^{\ast a}}_{\mu\nu}$, one finds that
%
%\begin{widetext}
%\begin{eqnarray}
%{R^\rho}_{\sigma\mu\nu} & = & {e_a}^\rho {e^b}_\sigma {R^a}_{b\mu\nu}=
%\partial_\mu{\Gamma^\rho}_{\sigma\nu}-\partial_\nu{\Gamma^\rho}_{\sigma\mu}
%+{\Gamma^\rho}_{\lambda\mu}{\Gamma^\lambda}_{\sigma\nu}-{\Gamma^\rho}_{\lambda\%nu}
%{\Gamma^\lambda}_{\sigma\mu} - H_{\mu\nu}\delta_\sigma^\rho,
%\label{wgtcurvature}\\
%{T^{\ast\lambda}}_{\mu\nu} & = & {e_a}^\lambda{T^{\ast a}}_{\mu\nu}
%= {\Gamma^\lambda}_{\nu\mu}- {\Gamma^\lambda}_{\mu\nu},
%\label{wgttorsion}\\
%H_{\mu\nu} & = & \partial_\mu B_\nu -\partial_\nu B_\mu.
%\label{wgthtensor}
%\end{eqnarray}
%\end{widetext}
%
%
\begin{eqnarray}
{R^\rho}_{\sigma\mu\nu} & = &
2(\partial_{[\mu}{\Gamma^\rho}_{|\sigma|\nu]}
+{\Gamma^\rho}_{\lambda[\mu}{\Gamma^\lambda}_{|\sigma|\nu]})- H_{\mu\nu}\delta_\sigma^\rho,\phantom{AAA}
\label{wgtcurvature}\\
H_{\mu\nu} & = & 2\partial_{[\mu} B_{\nu]},
\label{wgthtensor}\\
{T^{\ast\lambda}}_{\mu\nu}
& = & 2{\Gamma^\lambda}_{[\nu\mu]},
\label{wgttorsion}
\end{eqnarray}
where ${R^\rho}_{\sigma\mu\nu} = {e_a}^\rho {e^b}_\sigma
{R^a}_{b\mu\nu}$ and ${T^{\ast\lambda}}_{\mu\nu} =
{e_a}^\lambda{T^{\ast a}}_{\mu\nu}$. Thus, although we recognise
${T^{\ast\lambda}}_{\mu\nu}$ as (minus) the torsion tensor of the
$Y_4$ spacetime, we see that ${R^\rho}_{\sigma\mu\nu}$ is not simply
its Riemann tensor. Rather, the Riemann tensor of the $Y_4$ spacetime
is given by
\begin{equation}
{\widetilde{R}^{\rho}}_{\phantom{\rho}\sigma\mu\nu} \equiv {R^\rho}_{\sigma\mu\nu} +
H_{\mu\nu}\delta_\sigma^\rho.
\label{wgtrtildedef}
\end{equation}
One should note that, although $\widetilde{R}_{\rho\sigma\mu\nu}$ is
antisymmetric in $(\mu,\nu)$, it is no longer antisymmetric in
$(\rho,\sigma)$ (indeed $\widetilde{R}_{(\rho\sigma)\mu\nu} =
g_{\rho\sigma}H_{\mu\nu}$) and does not satisfy the familiar cyclic and
Bianchi identities of the Riemann tensor in a Riemannian $V_4$
spacetime.  One may also show that, with the given arrangements of
indices, both ${\widetilde{R}^{\rho}}_{\phantom{\rho}\sigma\mu\nu}$
(or ${R^{\rho}}_{\sigma\mu\nu}$) and ${T^{\ast\lambda}}_{\mu\nu}$
transform covariantly with weight $w=0$ under a local dilation.  It is
also worth noting that $\widetilde{R}_{\mu\nu} \equiv
{\widetilde{R}_{\mu\lambda\nu}}^{\phantom{\mu\lambda\nu}\lambda}
= R_{\mu\nu}- H_{\mu\nu}$ and $\widetilde{R} \equiv
{\widetilde{R}^{\mu}}_{\phantom{\mu}\mu} = R$. As one might expect,
the quantities (\ref{wgtcurvature})--(\ref{wgtrtildedef}) arise
naturally in the expression for the commutator of two derivative
operators acting on a vector $V^\rho$ (say) of Weyl weight $w$, which
is given by
\begin{equation}
[\nabla^\ast_\mu,\nabla^\ast_\nu] V^\rho =
{\widetilde{R}^{\rho}}_{\phantom{\rho}\sigma\mu\nu}V^\sigma +
wH_{\mu\nu} V^\rho - {T^{\ast\sigma}}_{\mu\nu}\nabla^\ast_\sigma V^\rho.
\label{wgtnablacomm}
\end{equation}
%

%%%%%%%%%%%%%%%%%%%%%%%%%%%%%%%%%%%%%%%%%%%%%%%%%%%%%%%%%%%%%%%%%%%%%%%%%%
%\subsection{Weyl covariant derivative}
%\label{sec:wcd}

A key point to note in the above geometric interpretation is that it
leads to the identification of a covariant derivative $\nabla^\ast_\mu
= \partial^\ast_\mu +
{\Gamma^\sigma}_{\rho\mu}{\matri{X}^\rho}_\sigma$ (often termed the
scale covariant or Weyl covariant derivative, although it was first
introduced by Dirac \cite{Dirac73}, who called it the co-covariant
derivative) that clearly differs from the conventional covariant
derivative $\nabla_\mu = \partial_\mu +
{\Gamma^\sigma}_{\rho\mu}{\matri{X}^\rho}_\sigma$ in Weyl--Cartan
spacetimes, since $\nabla_\mu^\ast = \nabla_\mu + wB_\mu$.  Indeed,
this form immediately leads to the important property
$\nabla^\ast_\sigma g_{\mu\nu}=0$.  A further important feature of the
Weyl covariant derivative is that, when acting on an object of weight
$w$, the resulting object also transforms covariantly with weight $w$;
this is not the case for the conventional covariant derivative
$\nabla_\mu$, which does not, in general, produce an object that
transforms covariantly, unless $w=0$.

It is also noteworthy that the Weyl covariant derivative cannot, in
general, be written in the form $\nabla^\ast_\mu = \partial_\mu +
{^\ast{\Gamma^\sigma}}_{\rho\mu}{\matri{X}^\rho}_\sigma$ for some
alternative connection ${^\ast{\Gamma^\sigma}}_{\rho\mu}$, even if the
latter is permitted to depend on $w$. Indeed, this is a manifestation
of a larger issue. Whereas the geometric interpretation of PGT
captures all of its gravitational interactions (at least for tensor
fields) in terms of the metric and connection of an underlying
Riemann--Cartan $U_4$ spacetime in which the matter fields reside, the
geometric interpretation of WGT does not describe all of its
gravitational interactions in an analogous manner. In particular, when
acting on fields with non-zero Weyl weight $w$, the gravitational
interactions mediated by the dilational gauge field $B_\mu$ cannot be
fully `geometrized' in terms of the metric and connection of a
Weyl--Cartan $Y_4$ spacetime, as is clear from (\ref{eq:ptdef}) and
(\ref{wgtnablacomm}), and one must augment the $Y_4$ spacetime
interpretation by using the Weyl covariant derivative in such cases.

\begin{comment}
In a wider context, this contradicts the generally held view that, for
any given gravitational theory, in choosing to interpret some
quantities as geometric properties of the underlying spacetime and
others as fields residing in that spacetime, one can place the
dividing line anywhere, with the gauge theory approach and its
geometric interpretation representing extreme ends of a range of
possibilities: the former interprets all quantities as fields residing
in a background Minkowski spacetime, whereas the latter interprets all
gravitational quantities as geometric properties of the
underlying spacetime. For WGT, however, the wholesale geometric
interpretation is not possible in the presence of fields with non-zero
Weyl weight.
\end{comment}

%%%%%%%%%%%%%%%%%%%%%%%%%%%%%%%%%%%%%%%%%%%%%%%%%%%%%%%%%%%%%%%%%%%%%%%%%%
\section{Second clock effect in WGT}
\label{sec:sceinwgt}

We now reconsider the second clock effect in the context of the above
geometric interpretation of WGT, in particular making use of the Weyl
covariant derivative that it identifies. Following our discussion in
Section~\ref{sec:WCST}, we will consider the SCE both in terms of the
norms of parallel transported vectors and in terms of an appropriately
defined proper time.

\myNew{As discussed in the
  Introduction,} in the geometric interpretation of WGT, 
\myNew{for a} test particle moving along some timelike
worldline $\mathcal{C}$ given by $x^\mu=x^\mu(\lambda)$, the
components of the tangent vector to this worldline as measured
\myNew{in the local Lorentz frame of an}
observer will be $u^a(\lambda) = {e^a}_\mu
u^\mu(\lambda)$\myNew{, which should be invariant under Weyl gauge
  transformations since they are physical observables in WGT}.  Since
the vierbein ${e^a}_\mu$ has weight $w=1$, the weight of the
components \myNew{$u^\mu(\lambda)$ in the coordinate
  basis} is thus \myNew{$w=-1$}.

One may perform calculations in either the tetrad or coordinate basis,
denoted by Latin and Greek indices, respectively. By virtue of the
geometric interpretation of WGT described in Section~\ref{sec:geo},
these two approaches yield consistent results, but we will work in
terms of the coordinate basis to facilitate a more straightforward
comparison with the calculations performed in Section~\ref{sec:WCST}.

\subsection{Parallel transported vectors}

Defining parallel transport in terms of the Weyl covariant derivative,
as in (\ref{eq:ptdef}), and using the condition $\nabla^\ast_\sigma
g_{\mu\nu}=0$, one immediately finds that, in contrast to
(\ref{eq:normevol}), the evolution of the inner product of any two
vectors $v^\mu$ and $w^\mu$ parallel transported along some curve
$\mathcal{C}$ is now given by
\bea
\nd{}{\lambda}[g_{\mu\nu}v^\mu(\lambda)w^\nu(\lambda)] 
&=& \frac{D^\ast}{D\lambda}[g_{\mu\nu}v^\mu(\lambda)w^\nu(\lambda)],\nonumber\\
&=& (u^\sigma(\lambda)\nabla^\ast_\sigma g_{\mu\nu}) v^\mu(\lambda)
w^\nu(\lambda) =0.\phantom{AAa}
\label{eqn:dpcons}
\eea
Hence, by setting $v^\mu=w^\mu$ and considering parallel transport
around a closed curve ${\cal C}$, the length $\ell$ of a vector is
unchanged on completing a loop, and so the original basis for
suggesting the existence of a SCE is removed.

\begin{comment}
One can also verify this conclusion by working instead in the tetrad
basis, for which (\ref{eqn:dpcons}) is replaced by
%
\bea
\nd{}{\lambda}[\eta_{ab}v^a(\lambda)w^b(\lambda)] 
&=& \frac{D^\ast}{D\lambda}[\eta_{ab}v^a(\lambda)w^b(\lambda)],\nonumber\\
&=& (u^c(\lambda){\cal D}^\ast_c \eta_{ab}) v^a(\lambda) w^b(\lambda)\\
&=& (u^\sigma(\lambda)D^\ast_\sigma \eta_{ab}) v^a(\lambda) w^b(\lambda)
 =0,\phantom{AAa}
\label{eqn:dpconstetrad}
\eea
%
since $D^\ast v^a/D\lambda = 0$. Check:
%
\bea
\frac{D^\ast v^a}{D\lambda} 
& = & \frac{D^\ast ({e^a}_\mu v^\mu)}{D\lambda} \\
& = & (u^c {\cal D}^\ast_c {e^a}_\mu)v^\mu + {e^a}_\mu \frac{D^\ast v^\mu}{D\lambda} \\
& = & (u^\sigma \Delta^\ast_\sigma {e^a}_\mu)v^\mu =0.
\eea
%
\end{comment}

\subsection{Proper time}

As discussed in Section~\ref{sec:WCST}, however, the intuitive
argument above is not rigorous, and so we now reconsider how to define
a physically sensible notion of proper time along (timelike)
worldlines.

We begin by following an analogous procedure to that adopted in
Section~\ref{sec:propertime}.  In particular, we first seek to
identify a parameter $\xi$ (it will become clear shortly 
\myNew{that this cannot be interpreted as proper time and
so} we do not
denote this variable by $\tau$ here) that satisfies an analogous
condition to (\ref{eqn:clockdef}), namely
\be
g_{\mu\nu}u^\mu(\xi)a^{\ast\nu}(\xi) = 0,
\label{eqn:clockdef2}
\ee
where we define $a^{\ast\mu} = D^\ast u^\mu/D\xi$. It is
straightforward to show that the condition (\ref{eqn:clockdef2}) is
consistent across the whole equivalence class defined by the gauge
transformations (\ref{eqn:weylgt}), since
\be
\bar{g}_{\mu\nu}\bar{u}^\mu(\xi)\bar{a}^{\ast\nu}(\xi) = 
e^{-\rho}g_{\mu\nu}u^\mu(\xi)a^{\ast\nu}(\xi).
\label{eq:orthoginv}
\ee
Following through the calculations in Section~\ref{sec:propertime},
but working instead in terms of the Weyl covariant derivative, one
finds that (\ref{eqn:taucalc1}) is replaced by
\bea
\nd{}{\lambda}[g_{\mu\nu}u^\mu(\lambda)u^\nu(\lambda)] 
&=& \frac{D^\ast}{D\lambda}[g_{\mu\nu}u^\mu(\lambda)u^\nu(\lambda)],\nonumber\\
&=& (u^\sigma\nabla^\ast_\sigma g_{\mu\nu}) u^\mu u^\nu  + 2g_{\mu\nu}u^\mu
a^{\ast\nu},\nonumber\\
&=& 2g_{\mu\nu}u^\mu a^{\ast\nu},
\label{eq:constl}
\eea
where we used the condition (\ref{eqn:nablastar}), and in the last two
lines (and the remainder of this section) we drop the explicit
dependence of quantities on the arbitrary parameter $\lambda$ for
brevity. Thus, as might be expected, the condition
(\ref{eqn:clockdef2}) corresponds simply to finding a parameterisation
$\xi$ for which the length of the tangent vector remains constant
under parallel transport along its worldline, as in Riemann--Cartan
$U_4$ spacetime. Consequently, (\ref{eqn:deltataudef}) is replaced by
\be
\Delta\xi(\lambda) =
\left.\frac{d\xi/d\lambda}{\sqrt{g_{\mu\nu}u^\mu
    u^\nu}}\right|_{\lambda_0} 
\int_{\lambda_0}^\lambda \sqrt{g_{\mu\nu}u^\mu
    u^\nu}\,d\lambda',
\label{eqn:deltaxidef}
\ee
which now gives the \myNew{parameter} interval $\Delta\xi$ between two
events corresponding to the parameter values $\lambda_0$ and $\lambda$
along the worldline. As was the case in Section~\ref{sec:propertime},
if $\xi$ is a solution then so too is $a\xi + b$, where $a$ and $b$
are constants. Thus, without loss of generality, one may choose $\xi$
such that the length of the tangent vector
$g_{\mu\nu}u^\mu(\xi)u^\nu(\xi)$ is unity along the entire worldline,
so that $u^\mu(\xi)$ may be interpreted as the particle 4-velocity,
and hence identified with the timelike unit basis $\hat{\mathbf{e}}_0$
of a local Lorentz frame for an observer moving along the worldline.

\begin{comment}
Again, one can verify that the same results are obtained in the tetrad
basis. One begins with the analogue of (\ref{eqn:clockdef2}), namely
%
\be
\eta_{ab}u^a(\xi)a^{\ast b}(\xi) = 0,
\ee
% 
where we define $a^{\ast b} = {e^b}_\mu a^{\ast \mu} = {e^b}_\mu
D^\ast u^\mu/D\xi$. Then, one can easily show that
(\ref{eq:orthoginv}) is replaced by
%
\be
\bar{\eta}_{ab}\bar{u}^a(\xi)\bar{a}^{\ast b}(\xi) = 
%e^{-\rho}
\eta_{ab}u^a(\xi)a^{\ast b}(\xi).
\ee
%
Following through the calculations 
%in Section~\ref{sec:propertime},
but working instead in the tetrad basis, one finds that
(\ref{eq:constl}) is replaced by
%
\bea
\nd{}{\lambda}[\eta_{ab}u^a(\lambda)u^b(\lambda)] 
&=& \frac{D^\ast}{D\lambda}[\eta_{ab}u^a(\lambda)u^b(\lambda)],\nonumber\\
&=& (u^c{\cal D}^\ast_c \eta_{ab}) u^a u^b  + 2\eta_{ab}u^a a^{\ast b},\nonumber\\
&=& 2\eta_{ab}u^a a^{\ast b},
\eea
%
where we have used the two results
%
\be
u^c{\cal D}^\ast_c \eta_{ab} = u^\sigma(\lambda)D^\ast_\sigma
\eta_{ab} = 0
\ee
%
and
%
\bea
\frac{D^\ast u^a}{D\lambda} 
& = & \frac{D^\ast ({e^a}_\mu u^\mu)}{D\lambda} \\
& = & (u^c {\cal D}^\ast_c {e^a}_\mu)u^\mu + {e^a}_\mu \frac{D^\ast u^\mu}{D\lambda} \\
& = & (u^\sigma \Delta^\ast_\sigma {e^a}_\mu)u^\mu 
+ {e^a}_\mu a^{\ast\mu}  = a^{\ast a}.
\eea
%

\end{comment}

As we discussed in Section~\ref{sec:uweight}, however, the
differential $d/d\xi$ has Weyl weight $w=-1$ (indeed this holds for
any arbitrary parameterisation $\lambda$ of the worldline). Thus,
$\xi$ is not invariant under Weyl gauge transformations, and so cannot
be interpreted as the proper time of a particle (or observer) moving
along the worldline, which is a physical observable and hence should
be independent of any gauge transformations.

To address this issue, one must recognise that Einstein's original
objection to Weyl's theory requires a massive Dirac field to represent
atoms and observers, and also take seriously the physical mechanism by
which such an observer might measure their proper time as they move
along their worldline. One such method would be to carry with them
some form of atomic clock, which provides a good physical
approximation to an ideal clock, and is used to define the standard
for the unit of time. The functioning of such a clock is based on the
spacing of energy levels in atoms (this is, of course, also directly
relevant to the consideration of spectral lines, the sharp nature of
which is considered as the key observational evidence against the
existence of the SCE).  Although not particularly practical, one could
in principle make use of the energy levels in the hydrogen atom, the
spacings of which are characterised in the clock's local Lorentz frame
by the Rydberg (ground-state to free) energy $E_{\rm R} = \tfrac{1}{2}
m \alpha^2$ (in natural units), where $m$ is the rest mass of the
electron and $\alpha$ is the (dimensionless) fine structure constant.

As pointed out in Section~\ref{sec:wgt}, however, to incorporate a
Dirac field describing `ordinary' matter  in WGT one must also introduce
a scalar compensator field $\phi$ and make the replacement
$m\bar{\psi}\psi \to \mu\phi\bar{\psi}\psi$ in the Dirac action, where
$\mu$ is a dimensionless parameter but $\mu\phi$ has the dimensions of
mass in natural units. Thus, the Rydberg energy \myNew{then} becomes
$E_{\rm R} = \tfrac{1}{2} \mu\phi\alpha^2$, and so in general varies
with spacetime position according to the \myNew{value} of
$\phi$. \myNew{A photon emitted in a ground-state to free electronic
  transition has energy $E_{\rm R}$, defined as the projection of the
  photon 4-momentum onto the timelike basis vector
  $\hat{\mathbf{e}}_0$ of the atom's local Lorentz frame, such that
  $E_{\rm R} = p_\mu dx^\mu/d\xi$. Therefore, in a small parameter
  interval $d\xi$, the number of cycles traversed in the photon wave
  train is $dN \propto E_{\rm R}\,d\xi \propto \phi\,d\xi$, which is
  invariant under a Weyl gauge transformation, as it must be, since
  $\phi$ and $d\xi$ have weights $w=-1$ and $w=1$, respectively.
   A proper time interval
  $d\tau$ in the atom's rest frame is, however, {\it defined} as the
  duration of a given number of cycles, and so $d\tau \propto
  \phi\,d\xi$, where one can take the constant of proportionality to
  equal unity, without loss of generality.} Hence the proper time
interval measured by \myNew{an atomic} clock between two events
corresponding to the parameter values $\xi_0$ and $\xi$ along the
worldline is given simply by
\be
\Delta\tau(\xi) = \int_{\xi_0}^\xi \phi\,d\xi',
\label{eqn:ptdeffinal}
\ee
\myNew{which} is invariant under Weyl gauge transformations, as required
for a physically observable quantity. \myNew{Indeed, the Rydberg energy
$E_{\rm R}$ is then equal (in natural units) to the angular frequency of
the photon as measured in terms of the proper time $\tau$, and is
itself also invariant under Weyl gauge transformations, as it should
be.}

Finally, applying this proper time definition to the two-clock thought
experiment discussed in the Introduction, one sees immediately from
(\ref{eqn:deltaxidef}) and (\ref{eqn:ptdeffinal}) that the ratio of
the elapsed proper time measured by each clock between their reunion
and some subsequent event along their joint worldline is unity. Thus,
the clock rates are the same after their reunion, and so WGTs do not
predict a SCE.

\begin{comment}
In the sort of device you mention, we need to know how a 'tick' of the
clock would be registered. With a photon bouncing backwards and forwards
in a cavity it is not obvious how we physically determine when a photon
'bounces' off a wall/mirror. We *can* determine frequencies where there
are resonant effects of the EM field, but this would bring us back to
things like hydrogen masers, and spectral lines, where the compensator
field does matter. So what do you see as being the actual mechanism for
measuring time with such a device? From that one could seek to find out
how the measurement could be affected by a scale gauge change.
\end{comment}

\medskip
%%%%%%%%%%%%%%%%%%%%%%%%%%%%%%%%%%%%%%%%%%%%%%%%%%%%%%%%%%%%%%%%%%%%%%%%%%
\section{Conclusions}
\label{sec:conc}

We have critically reconsidered the prevailing view in the literature
that Weyl gauge theories of gravity (WGTs) predict a second clock
effect (SCE), which has previously been argued to rule out such
theories as unphysical. Although WGTs are interpreted geometrically in
terms of a Weyl--Cartan $Y_4$ spacetime, the gravitational
interactions mediated by the dilational gauge field (or Weyl potential)
$B_\mu$ cannot be fully `geometrized' in terms of the metric and
connection of such a spacetime when acting on quantities with non-zero
scaling dimensions (or Weyl weight) $w$. 

Rather, the geometric interpretation of WGTs identifies a covariant
derivative $\nabla_\mu^\ast$ (often termed the Weyl covariant
derivative) that differs from the conventional covariant derivative
$\nabla_\mu$ in Weyl--Cartan spacetime when acting on quantities of
non-zero Weyl weight. The Weyl covariant derivative has the important
properties that $\nabla_\sigma^\ast g_{\mu\nu}=0$ and, when acting on
quantities that transform covariantly with arbitrary weight $w$ under
Weyl gauge transformations, it produces objects that also transform in
this way; neither of these properties is shared by $\nabla_\mu$.  If
one defines parallel transport in terms of the Weyl covariant
derivative, then the condition $\nabla_\sigma^\ast g_{\mu\nu}=0$
immediately implies that the inner product of any two
vectors is preserved as they are parallel transported along some
curve, which removes the basis for Einstein's original concerns
regarding the existence of a SCE. 

Moreover, we show that more recent derivations of the SCE, which
are based on defining proper time in Weyl--Cartan spacetime, rely on
the assumption that the components $u^\mu$ of the tangent vector to an
observer's worldline are invariant ($w=0$) under Weyl gauge
transformations, whereas we show that, in fact, they have weight
$w=-1$.

Furthermore, we point out that Einstein's original objection to Weyl's
theory requires the presence of a massive Dirac matter field to
represent atoms, observers and clocks, so it is meaningless in this context to
consider an empty Weyl--Cartan spacetime. The requirement of a Dirac
field to represent such `ordinary' matter in turn necessitates the
inclusion of a scalar compensator field in order that the total action
obeys local Weyl invariance and the Dirac field may acquire a mass
dynamically. We show that this scalar field is key to a physically
meaningful definition of proper time.

When one makes use of the Weyl covariant derivative to define
variation along a worldline, assigns the components $u^\mu$ of the
tangent vector to an observer's worldline the correct Weyl weight of
$w=-1$, and includes the effect of the required compensator field in
defining a physically sensible proper time variable that is invariant
under Weyl gauge transformations, one immediately concludes that WGTs
do not predict a SCE.

%%%%%%%%%%%%%%%%%%%%%%%%%%%%%%%%%%%%%%%%%%%%%%%%%%%%%%%%%%%%%%%%%%%%%%%%%%%%%
%\begin{acknowledgments}
%We thank Will Barker for useful comments.
%\end{acknowledgments}
%%%%%%%%%%%%%%%%%%%%%%%%%%%%%%%%%%%%%%%%%%%%%%%%%%%%%%%%%%%%%%%%%%%%%%%%%%

\end{document}